\documentclass[aps,prl,twocolumn,superscriptaddress,preprintnumbers]{revtex4}
\usepackage{epsfig}
\usepackage{graphicx}
\usepackage{multirow} 
\usepackage{color}
\usepackage[normalem]{ulem}
\usepackage{nico} 
\usepackage{hyperref}

\newcommand\dublin{School of Mathematics, Trinity College, Dublin 2, Ireland}
\newcommand\edinb{SUPA, School of Physics, The University of
  Edinburgh, Edinburgh EH9 3JZ, UK}

\begin{document}

\preprint{Edinburgh 2012/12}
\preprint{TCDMATH 12-06}

\title{Neutral kaon mixing beyond the standard model with $n_f=2+1$ chiral
fermions }

\author{P.~A.~Boyle}
\affiliation{\edinb}
\author{N.~Garron}
\affiliation{\edinb} \affiliation{\dublin}
\author{R.~J.~Hudspith}
\affiliation{\edinb}

\collaboration{The RBC and UKQCD Collaborations}

\date{June 25th 2012}

\begin{abstract}  
We compute the hadronic matrix elements of the four-quark operators needed 
for the study of $K^0-{\bar K^0}$ mixing beyond the Standard Model (SM). 
We use $n_f=2+1$ flavours of domain-wall fermion (DWF) which exhibit
good chiral-flavour symmetry. 
The renormalization is performed
non-perturbatively through the RI-MOM scheme 
and our results are converted perturbatively to $\msbar$. 
The computation is performed on a single lattice spacing 
$a\sim0.086\, \fm $ with a lightest unitary pion mass of $290\, \MeV$. 
The various systematic errors, including the discretisation 
effects, are estimated and discussed. Our results confirm a 
previous quenched study, where large ratios of non-SM to SM matrix elements 
were obtained.
\end{abstract}

%\pacs{11.15.Ha, % Lattice gauge theory 
%11.30.Er	%Charge conjugation, parity, time reversal, and other discrete symmetries
%12.38.Gc  % Lattice QCD calculations
%   13.25.Es	%Decays of K mesons
%}

\maketitle

%\section*{Introduction}\label{sec:intro}
{\em Introduction.---}
Recent progress achieved by the lattice community 
is greatly improving our theoretical understanding of CP violation in 
kaon decays. The experimentally well-measured parameters $\varepsilon_K$
and $\varepsilon_K'$, which quantify indirect and direct CP violation 
in $K\to\pi\pi$ can be confronted with a theoretical computation
of the $K\to(\pi\pi)_{\rm I=0,2}$ amplitudes and of neutral kaon mixing,
providing the non-perturbative effects are correctly accounted for.
%The direct computation of $K\to\pi\pi$ decays 
%present many difficulties (both technical and theoretical), 
%and has been a real challenge, in particular for the lattice community.
%It is only recently that 
The first direct and realistic computation of the $K\to(\pi\pi)_{\rm I=2}$ 
amplitude has been recently performed~\cite{Blum:2011ng,Blum:2012uk}.
A complete computation of the $\Delta I=1/2$ amplitude is still missing, 
but important work is being performed in that direction~\cite{Blum:2011pu}.
The situation is much more favourable for neutral 
kaon oscillations: the bag parameter $B_K$ 
which describes the long-distance contributions
to neutral kaon mixing in the SM, is now computed with a precision 
of a few percent~(see e.g.~\cite{Aoki:2010pe, Durr:2011ap, Constantinou:2010qv})
for recent unquenched determinations).
By combining $B_K$ with the experimental value of $\varepsilon_K$,
one obtains important constraints on 
the free parameters associated with quark flavour mixing
(see~\cite{Lellouch:2011qw} for a recent pedagogical review).  
In principle, the same techniques can be applied for beyond the Standard Model (BSM) theories
(see for example~\cite{Bona:2007vi,Duling:2009pj,Enkhbat:2011qz,Buras:2010mh,Maiezza:2010ic,Buras:2010pz,Blanke:2011ry,Archer:2011bk,Buras:2011wi,Bao:2012nf}), 
but not much is known concerning the long-distance contributions 
of the non-standard operators beyond the quenched approximation.

In this letter, we present the first realistic computation 
of the matrix elements of neutral kaon mixing beyond the standard model.
Previous studies were either preliminary 
\cite{Wennekers:2008sg,Dimopoulos:2010wq}
or suffered from the quenched approximation
\cite{Donini:1999nn,Babich:2006bh}. 
Since a noticeable disagreement was
observed between 
\cite{Donini:1999nn} and \cite{Babich:2006bh}
it is important to repeat this computation in a more realistic
framework
~\footnote{The European twisted mass collaboration has recently reported on 
a similar study \cite{Bertone:2012cu}}
.\\ 
%--------------------------------------------------------------------------------
%\section*{Formalism}\label{formalism}

\begin{figure}[!t]
\begin{center}
\includegraphics[width=5cm]{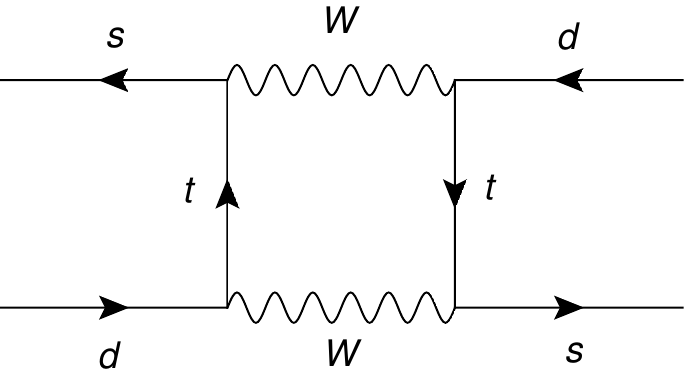}
\caption[]{%Example of a 
%Box 
Diagram contributing to $K^{0}-\overline{K^{0}}$
mixing in the SM. 
\vspace{-0.5cm}
} 
\label{fig:box}
\end{center}
\end{figure}

{\em Formalism.---}
In the SM, neutral kaon mixing is dominated 
by box diagrams as in figure~\ref{fig:box}.
By performing an operator product expansion, one factorises
the long-distance effects in 
the weak matrix element (WME)
$\langle \overline K^0 | O^{\Delta s=2}_1 | K^0 \rangle $ 
of the four-quark operator $O_1^{\Delta s=2}$ given by 
($\alpha$ and $\beta$ are colour indices)
\be
\label{eq:O1}
O_1^{\Delta s=2}=
(\overline s_\alpha \gamma_\mu (1-\gamma_5) d_\alpha)\,
(\overline s_\beta  \gamma_\mu (1-\gamma_5) d_\beta) \, .
\ee %,
Only one four-quark operator appears in the SM because neutral kaon
mixing occurs under W-boson exchange, 
implying a ``vector-axial'' Dirac structure. Since this four-quark operator is
invariant under Fierz re-arrangement, the two different colour structures (mixed and
unmixed) are equivalent.

We also seek to understand whether new physics beyond the
standard model could play a detectable role in kaon CP violation. 
Assuming that this new physics is perturbative, we might also describe this
by contributions to the effective Hamiltonian, 
and other four quark operators are induced by such extensions of the SM.
It is conventional to introduce the so-called SUSY basis
$O^{\Delta s =2}_{i=1\ldots5}$:
in addition to the SM operator $O_1^{\Delta s =2}$, we define~\cite{Gabbiani:1996hi}
\footnote{We discard the parity odd operators
since they are not relevant in the present case.}
\bea
\label{eq:OBSM}
O_2^{\Delta s=2}&=&
(\overline s_\alpha (1-\gamma_5) d_\alpha)\,
(\overline s_\beta  (1-\gamma_5) d_\beta)\nonumber,\\
\label{eqO3}
O_3^{\Delta s=2}&=&
(\overline s_\alpha  (1-\gamma_5) d_\beta)\,
(\overline s_\beta   (1-\gamma_5) d_\alpha)\nonumber,\\
O_4^{\Delta s=2}&=&
(\overline s_\alpha  (1-\gamma_5) d_\alpha)\,
(\overline s_\beta   (1+\gamma_5) d_\beta)\nonumber,\\
\label{eqO5}
O_5^{\Delta s=2}&=&
(\overline s_\alpha  (1-\gamma_5) d_\beta)\,
(\overline s_\beta   (1+\gamma_5) d_\alpha).
\eea
%---
These four-quark operators appear in the generic $\Delta s=2$ Hamiltonian
\be
\label{eq:H}
H^{\Delta s=2} = \sum_{i=1}^{5} \, C_i(\mu) \, O_i^{\Delta s=2}(\mu) \,,
\ee
where $\mu$ is a renormalization scale.
The Wilson coefficients $C_i$, which encode the short-distance effects, 
depend on the new physics model under consideration. The long-distance effects
are factorised into the matrix elements of the four-quark operators $O_i^{\Delta s=2}$. 
Lattice QCD offers a unique opportunity 
to quantify these effects in a model-independent way. 

In phenomenological applications, by combining our results for the BSM matrix
elements with experimental observables (typically the mass difference
$\Delta {M_K}= m_{K_L}-m_{K_S}$ and 
$\epsilon_K$), 
one obtains important constraints on the model under consideration.
The Wilson coefficients and the bare matrix elements
have to be converted into a common scheme, at a given scale $\mu$ (typically
$\rm \overline{MS}$ at 
a scale of $2$ or $3$ GeV). 
In the SM case ($i=1$) it is conventional to introduce the kaon bag
parameter $B_K$, which measures the deviation of the SM matrix element from its
vacuum saturation approximation (VSA)
\be
B_K = {  \langle \Kb| O_1 | \K \rangle  \over { {8\over3} m_K^2 f_K^2 }} \;
\;.
\ee
Where the normalisation for the decay constant is such that
$f_{K^-}\;=\:156.1 \MeV$. 
Several normalisations for the BSM operators have been proposed in the literature
(see for example~\cite{Donini:1999nn}), 
in this work we follow~\cite{Babich:2006bh} 
and define the ratios
\be
\label{eq:R}
R_i^{\rm BSM}  = 
\left[ {f_K^2 \over m_K^2} \right]_{\rm expt}
\left[ {m_P^2 \over f_P^2} { \langle \Pb| O_i | \P \rangle \over \langle \Pb|
O_1 | \P \rangle }\right]_{\rm latt} \;,
\ee
where $P$ is a pseudo-scalar particle of mass $m_P$ and decay constant $f_P$.
The term $\left[\:\right]_{\rm latt}$ is obtained from our lattice simulations
for different values of  $m_P$.

As advocated in~\cite{Babich:2006bh}, there are various reasons to choose this 
normalisation rather than, for
example, the standard bag parameters. In particular, we expect some systematic
errors to cancel in the ratio of the bare WMEs, there is no need to introduce
the light quark masses and the partial conservation of the axial current.
Furthermore, the mass factors in eq.(\ref{eq:R}) compensate the leading-order
chiral behaviour of the WMEs, making all of the chiral extrapolations smoother.
Finally, at the physical point $P=\K$, the $R_i^{\rm BSM}$'s give directly the ratios 
of the non-SM to SM contributions. 

The $R_i^{\rm BSM}$'s will be the main result of this work, but for completeness
we will also give the BSM bag parametrisation, defined as
(where $N_{2,\ldots,5}={\frac 53,-\frac 13, -2, -\frac 23} $)
\cite{Conti:1998ys},
\begin{equation}\label{eq::BSM_BAG}
B_i \;=\; -\frac{\langle \Kb | O_i | \K \rangle}{N_i \langle \Kb |
\bar{s}\gamma_5 d |0\rangle \langle 0 | \bar{s}\gamma_5 d| \K
\rangle},
\quad i=2,\ldots,5
.
\end{equation}

%--------------------------------------------------------------------------------
%\section*{Computation details}\label{sec:strategy}
{\em Computation details.---}
This computation is performed on the finer of the two ensembles described in detail
in~\cite{Aoki:2010pe,Aoki:2010dy}.
We use $32^3\times64\times16$ (the last number corresponds to the length
of the fifth dimension of the domain-wall action in lattice units)
Iwasaki gauge configurations with an inverse lattice spacing 
$a^{-1}=2.28(3)\,\GeV$, corresponding to $a\sim0.086\,\fm$. 
The quarks are described by the domain-wall
action~\cite{Kaplan:1992bt,Shamir:1993zy,Furman:1994ky}, 
both in the valence and in the sea sectors. We have three different values of
the 
light sea quark mass $am_{\rm light}^{\rm sea}=0.004, 0.006, 0.008$
corresponding to unitary pion masses
of approximately $290, 340$ and $390 \;\MeV$ respectively. 
The simulated strange sea quark mass is $am_{\rm strange}^{\rm sea}=0.03$,
which is close to its physical value $0.0273(7)$~\cite{Aoki:2010dy}. 
For the main results of this work, we consider only unitary light quarks, 
$am_{\rm light}^{\rm valence}=am_{\rm light}^{\rm sea}$, 
whereas for the physical strange we interpolate between the unitary 
($am_{\rm strange}^{\rm valence}= am_{strange}^{\rm sea}=0.03$) and the partially
quenched $(am_{\rm strange}^{\rm valence}=0.025) $ data. 

\begin{figure}[!h]
\begin{center}
\includegraphics[width=8cm]{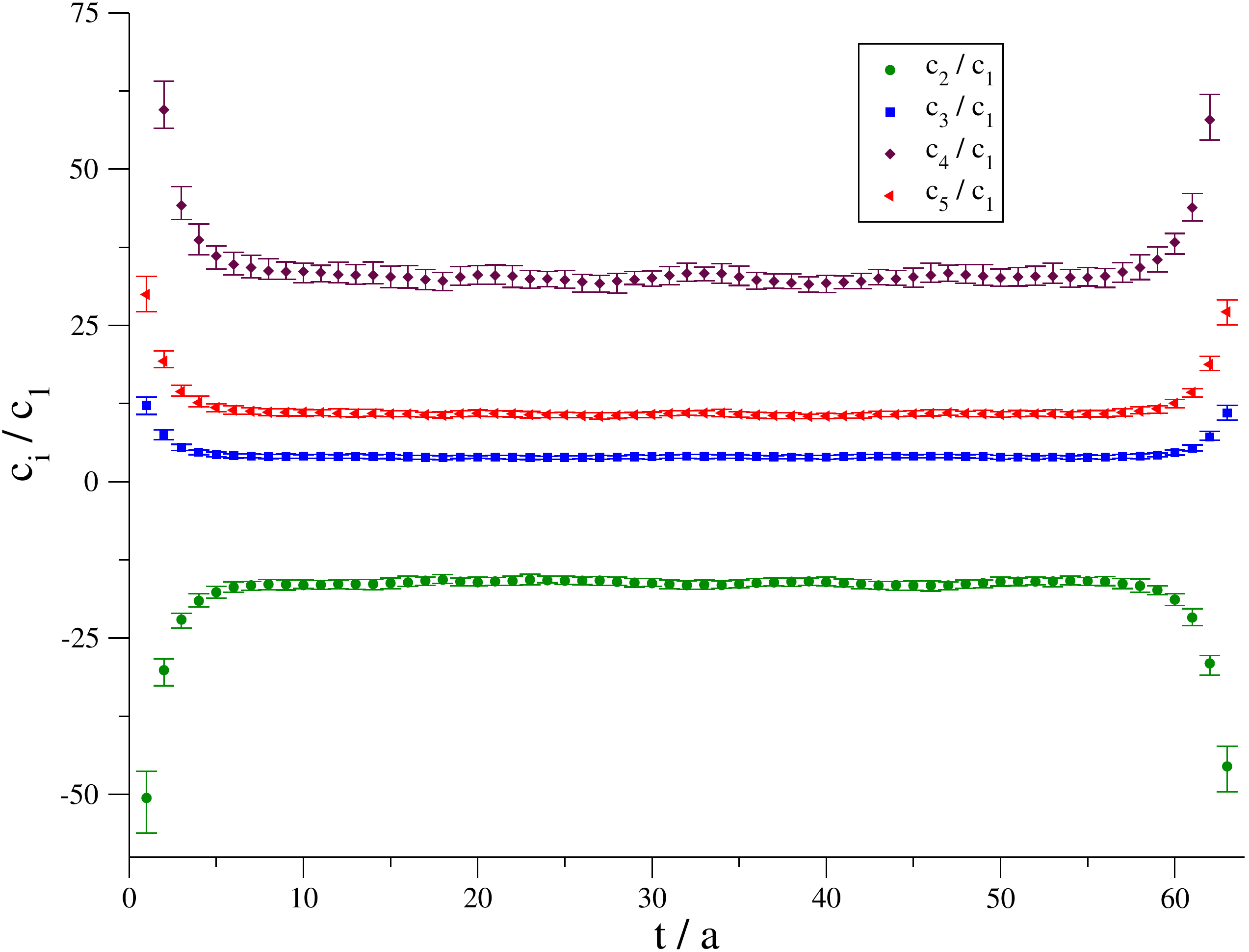}
\caption[]{Ratios of the bare three point functions from
which we extract $R_i^{\rm BSM}$.
Results are shown for our lightest simulated unitary kaon.
\vspace{-0.5cm}}
\label{fig:plateaux}
\end{center}
\end{figure}

To extract the bare matrix elements we follow ~\cite{Aoki:2010dy,Aoki:2010pe},
where the procedure for the evaluation of the two-point functions and of the three point function 
of the SM operator has been explained in detail.
In particular, we have used Coulomb gauge fixed wall sources to obtain very good
statistical precision.
From the measurement of $B_K$,
the generalisation to the BSM operators is straightforward.
We define the
three point function 
\begin{equation}
c_i = \la \Pb(t_f) O^{\Delta S=2}_{i=2,\ldots5}(t) \Pb(t_i) \ra.
\end{equation}
From the asymptotic Euclidean time behaviour 
of the ratios of three point-functions $c_i/c_1$
(we fit these ratios to a constant in the interval $t/a=[12,52]$ \footnote{From
figure~\ref{fig:plateaux} we deduce that we have reached the asymptotic region 
in this range for each operator insertion.}) we 
obtain the bare matrix elements of the four-quark operators 
normalised by the SM one:
$\left[ \langle \Pb| O_i^{\Delta S=2} | \P \rangle / \langle \Pb| O_1^{\Delta S=2} | \P \rangle\right]^{\rm bare}$ .
In figure~\ref{fig:plateaux}, we show the corresponding plateaux 
obtained for our lightest unitary kaon 
$am_{\rm light}^{\rm sea} =  am_{\rm light}^{\rm valence} =0.004$,
$am_{\rm strange}^{\rm sea} = am_{\rm strange}^{\rm valence}=0.03$.
\\
%--------------------------------------------------------------------------------
%\section*{Renormalization}\label{sec:renormalization}

{\em Renormalization.---}
The four-quark operators given in eq.~(\ref{eq:O1}-\ref{eq:OBSM}) mix under
renormalization. In a scheme which preserves chiral symmetry, the mixing pattern is given by the 
transformation properties of these operators under chiral rotations 
$SU(3)_L\times SU(3)_R$. 
The SM operator 
$O_1^{\Delta s=2}$ belongs to a $(27,1)$ irreducible representation of 
$SU(3)_L \times SU(3)_R$
and renormalizes multiplicatively.
The BSM operators fall into two categories: 
$O_2^{\Delta s=2}$ and $O_3^{\Delta s=2}$ transform like 
$(6,\overline 6)$ 
and mix together.
Likewise $O_4^{\Delta s=2}$ and $O_5^{\Delta s=2}$ belong to $(8,8)$
and also mix. If chiral symmetry is realised, 
the five-by-five renormalization matrix  $Z_{ij}$ is block diagonal:  
a single factor for the $(27,1)$
operator and two $2\times 2$ matrices for the BSM operators.
Because we work with the domain-wall fermions formulation,
in which the explicit breaking of chiral symmetry
can be made arbitrary small (and in practice numerically irrelevant),
we expect to find this continuum pattern, up to small discretisation effects.

\begin{figure}[!t]
\begin{center}
\includegraphics[width=8cm]{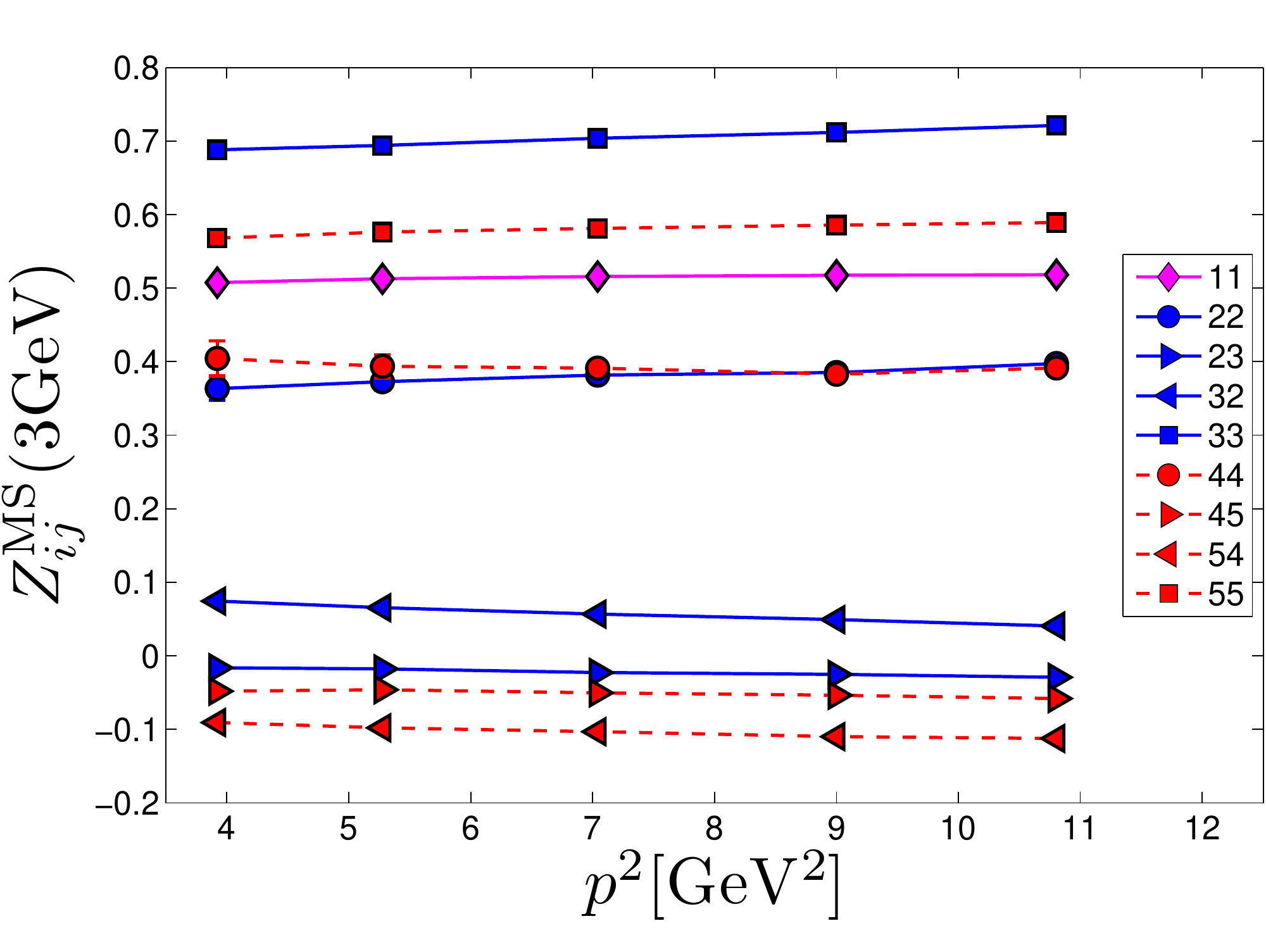}
\caption[]{Renormalization factors of the four-quark operators.
At each value of the simulated momentum $p$, we run to the scale of $3\GeV$ and 
convert to $\msbar$. The remaining scale dependence 
can be imputed to the truncation of the perturbative expansion.
We show only the $Z$-factors allowed by chiral symmetry.
\vspace{-0.5cm}
}
\label{fig:Z}
\end{center}
\end{figure}

\begin{figure}[t]
\begin{center}
\includegraphics[width=9cm]{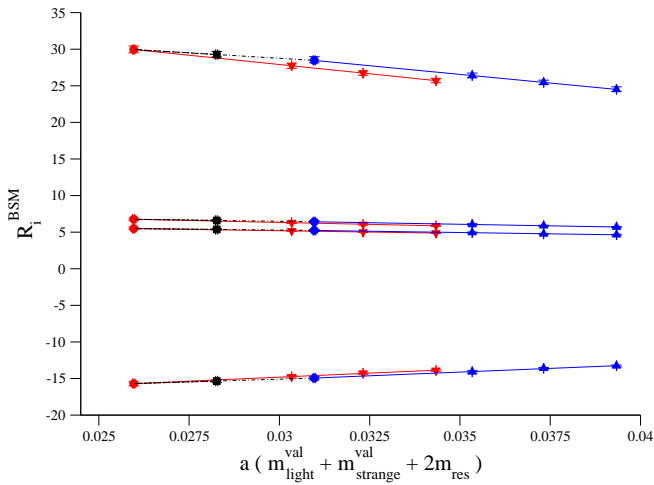}
\caption[]{Renormalized BSM ratios $R^{\rm BSM}_i$ in function of the bare 
valence quark mass.
We show the three unitary light quarks for both 
the unitary strange (upward blue triangles) and the partially
quenched strange (downward blue triangles), 
together with their extrapolations 
in the light sector (blue and red circles) and their interpolation 
to the physical kaon mass (black squares).
\vspace{-0.5cm}
}
\label{fig:extrap}
\end{center}
\end{figure}

We perform the renormalization of the four-quark operators $O_i^{\Delta s=2}$
non-perturbatively in the RI-MOM scheme~\cite{Martinelli:1994ty}.
We compute the forward,
amputated, vertex functions of the relevant operators between external quark states in 
the lattice Landau gauge for a given set of momenta. 
As a renormalization condition, we impose these to be equal to their tree-level
values once projected onto their colour-Dirac structures and extrapolated
to the chiral limit.
By using both momentum sources~\cite{Gockeler:1998ye} and partially
twisted boundary conditions, we obtain 
smooth functions of the external momentum with 
very good statistical accuracy~\footnote{More details about the computation of the
renormalization factors can 
be found in~\cite{Aoki:2010pe,Boyle:2011kn}}. 
Although we have performed this computation in various
intermediate schemes (including the non-exceptional
$(\gamma_\mu,\gamma_\mu)$-scheme introduced in~\cite{Aoki:2010pe})
we quote here the results obtained via the RI-MOM scheme
because only in this case are the conversion factors to $\msbar$
(computed in continuum perturbation theory) available for the whole set of operators.
In such a scheme, 
the presence of exceptional channels enhances the Goldstone pole contaminations
\cite{Aoki:2007xm,Wennekers:2008sg}, 
which have then been subtracted explicitly~\cite{Giusti:2000jr}.
We choose to impose the renormalization conditions at 
$\mu=3\,\GeV$. 
At this scale, perturbation theory (PT) converges rather quickly ($\alpha_s\sim0.25$),
the chiral symmetry breaking effects are small, 
and we still have good control on the discretisation 
effects~\cite{Arthur:2011cn,Aoki:2010pe}.
Matching at this scale is important for our error budget since the perturbative 
conversion factors between the lattice schemes and $\msbar$ of the four-quark operators 
are only known at one-loop (as discussed later, matching to PT is actually one of our 
dominant sources of systematic error).
We observe that the effects of chiral symmetry breaking are not 
completely negligible, even at $3\,\GeV$~\cite{Boyle:2011kn}.
Therefore we must assess a systematic error to the mixing of operators
of different chirality (see next section).
We have checked that in a non-exceptional scheme this small 
chirally forbidden mixing is strongly reduced and becomes numerically irrelevant 
at $3\,\GeV$~\cite{Boyle:2011kn,Wennekers:2008sg}. Thus we conclude 
that this effect is a physical manifestation of the infrared behaviour of the exceptional 
intermediate scheme. 
The results for the chirally allowed renormalization factors $Z_{ij}^{\msbar}(3\GeV)$
are shown in figure~\ref{fig:Z}.
They relate the bare four-quark operators to the 
renormalized ones through the usual relation 
($Z_q$ is the renormalization factor of the quark wave 
function and cancels in the ratios)
\be
O_i^{\Delta s=2,\rm \msbar}({3\GeV}) = {Z_{ij}\over Z_q^2}^{\msbar}\!\!\!\!({3\GeV})\,
{O_j^{\Delta s=2,\rm bare}}
\ee
%--------------------------------------------------------------------------------
%\section*{Physical results and error estimation}\label{sec::Results}
{\em Physical results and error estimation.---}
Once the bare ratios have been renormalized, we extrapolate them to the 
physical kaon mass. 
We have explored different strategies, such as a simple polynomial form
or next-to-leading order chiral perturbation theory (ChPT) 
predictions~\footnote{The functional forms of the B parameters predicted by ChPT 
can be found in ~\cite{Becirevic:2004qd},
%and has also been recently recomputed 
by staggered ChPT in~\cite{Bailey:2012wb},
and by heavy-meson ChPT in~\cite{Detmold:2006gh}}.
%We have also performed the extrapolations 
%using some lighter partially quenched data points obtained 
%from these ensembles. 
The value of the simulated unitary 
strange mass differs somewhat from the physical one, therefore we perform 
an interpolation using a partially quenched strange quark.
We find that the $R^{\rm BSM}$'s exhibit a very mild quark mass dependence
(see figure~\ref{fig:extrap}~\footnote{Because of a small explicit chiral symmetry 
breaking due to the finite fifth dimension
of the domain-wall action, a residual mass $m_{\rm res}$ has to be added to the 
bare quark mass; on this lattice $am_{\rm res}\sim 10^{-4}$~\cite{Aoki:2010dy}.}), 
therefore we take the results 
obtained with the polynomial Ansatz %for the unitary light points
as our central values~\footnote{Our strategy to obtain the values of the bare quark masses 
at the physical point differs slightly from the one 
presented in ~\cite{Aoki:2010dy} but we find compatible results.}.

Our final results 
are the $R_i^{\rm BSM}$ quoted in $\msbar$ at $3\, \GeV$ 
given in table \ref{tab:results}. 
For completeness, we also convert these to the 
BSM bag parameters, using eq.~(\ref{eq::BSM_BAG})~\footnote{The details of the analysis
of the two-point functions needed for the $B$-parameters can be found in~\cite{Aoki:2010dy}.}. 
We also note that, using the same framework, the SM contribution is found to 
be 
$B_1 = B_K = 0.517 (4)_{\rm stat}$
in the $\msbar$ scheme at $3\,\GeV$, 
whereas a continuum value of 
$0.529(5)_{\rm stat} (19)_{\rm syst}$ was quoted in~\cite{Aoki:2010pe}. 
The difference comes from the fact that a different intermediate 
scheme was used in~\cite{Aoki:2010pe} (such a difference is accounted for in
our estimation of the systematic errors). 
From the same reference, the discretisation 
effects for $B_K$ on this lattice are seen to be of the order of $1.5\%$.
Since we have only one lattice spacing for the BSM ratios,
we make the assumption that the discretisation errors are 
of the same size as those affecting $B_K$, and estimate a $1.5\%$ error
to all the different operators.
This might appear like a crude estimate, but this effect is expected to be 
sub-dominant compared to other sources of systematic errors.
The next systematic error (called ``extr.'') represents the spread 
of the results obtained from different extrapolation strategies
to the physical point.
The systematic error associated with the non-perturbative renormalization
(NPR) has been estimated from the breaking of chiral symmetry.
The mixing of the $(6,\bar 6)$ with the $(8,8)$ operators 
is forbidden by chiral symmetry, but likely to be enhanced 
by the exchange of light pseudo-scalar particles.
% This effect is in general of the order of $3-4\%$, 
%except for $O_2$ where this effect is $\sim 9\%$. 
%This is seen to come from the mixing
%between $O_2^{\rm \Delta s=2}$ and $O_4^{\rm \Delta s=2}$,
%which is forbidden by chiral symmetry, but likely to be enhanced 
%by the exchange of light pseudo-scalar particles.
As the matrix element of $O_4^{\rm \Delta s=2}$
is numerically large, this non-physical mixing has
effect on $O_2^{\rm \Delta s=2}$ and $O_3^{\rm \Delta s=2}$ 
of the order of $8-9\%$.
As discussed in the previous section,
this unwanted infrared effect is absent if a non-exceptional 
scheme is used.
%. 
The last error we quote (``PT'') arises from the matching 
between the intermediate RI-MOM scheme and $\msbar$,
which is performed at one-loop order in perturbation 
theory~\cite{Ciuchini:1993vr,Buras:2000if}
in the three-flavour theory.
The associated error is obtained by taking half the difference between the leading
order and the next to leading order result~\footnote{
To obtain $\alpha_s$ at $3\,\GeV$ in the three-flavour theory,
we start from $\alpha_s(M_Z) =0.1184$ 
\cite{Nakamura:2010zzi}, we use the four-loop 
running~\cite{vanRitbergen:1997va,Chetyrkin:1997sg} 
to compute the scale evolution down 
to the charm mass, while changing the number of flavours when crossing a threshold,
and then run up to $3\,\GeV$ in the three-flavour theory.}.
We note that this error is one of the dominant ones in our budget,
and we expect this error to be reduced by an important factor 
if a non-exceptional scheme were used, since the latter are known to converge faster
in perturbation theory. 
We neglect the finite volume effects which have been found to be small
in~\cite{Aoki:2010pe}, as one can expect from the value of $m_\pi L\sim 4$ 
for our lightest pion mass $m_\pi\sim 290\, \MeV$.
\begin{table}[t]
\begin{center}
\begin{tabular}{ c | c  c  ||  c  c  c  c  c | c }
\hline
i & $R_i^{\rm{BSM}}$ & $B_i$ 
& stat. & discr. & extr. & NPR & PT  & total
\\
\hline
$2$  & $-15.3 (1.7)$  & 0.43 (5)  & $1.3$  & $1.5$ & $4.0$  & $9.4$ & $4.7$   & $11.3$ \\
$3$  & $  5.4 (0.6)$  & 0.75 (9)  & $2.0$  & $1.5$ & $3.9$  & $7.8$ & $7.6$   & $12.0$  \\ 
$4$  & $ 29.3 (2.9)$  & 0.69 (7)  & $1.3$  & $1.5$ & $4.1$  & $3.0$ & $8.2$   & $9.8$  \\ 
$5$  & $  6.6 (0.9)$  & 0.47 (6)  & $2.1$  & $1.5$ & $3.8$  & $3.2$ & $12.6$  & $13.8$
\end{tabular}
\caption{Final results of this work: the first two columns 
show the ratios $R_i^{\rm BSM}$ and the corresponding bag
parameters $B_i$ in $\msbar$  at $3\,\GeV$,
together with their total error, combining systematics and statistics.  
In the remaining columns, we give our error budget 
for the $R^{\rm BSM}$, 
detailing the contributions in percentage of the different sources 
of systematics (see text for more details).
\vspace{-0.5cm}
}
\label{tab:results}
\end{center}
\end{table}

%
%--------------------------------------------------------------------------------
%\section*{Conclusions}\label{sec::Conclusions}
{\em Conclusions.---}
We have computed the electroweak matrix elements of the $\Delta s=2$ 
four-quark operators which contribute 
to neutral kaon mixing beyond the SM. 
Our work improves on the the previous studies by using 
both dynamical and chiral fermions.
We confirm the effect seen in a previous quenched 
computation \cite{Babich:2006bh}, 
where a large enhancement of the non-standard matrix elements were observed. 
The errors quoted in this work are of the order of $10\%$. 
We note that the main limitation of this study comes from the lack of matching factors between 
non-exceptional renormalization schemes (such as SMOM) and $\msbar$. 
Once these factors are
available, we expect to reach a precision better than $5\%$.
We also plan to utilise another lattice spacing 
in order to have a better handle on the discretisation effects.%\\

%--------------------------------------------------------------------------------
\section*{Appendix}
We have computed the non-perturbative scale evolution of the $R^{\rm BSM}$'s
between $3$ and $2\,\rm GeV$, and then converted the results to $\msbar$ 
using one-loop perturbation theory~\cite{Ciuchini:1993vr,Buras:2000if}:
\bea
&&
U^{\msbar}(2\,{\rm GeV}, 3\,{\rm GeV})
= \qquad \qquad\qquad \qquad  \nn\\
&& 
\left(
\begin{array}{r r r r r }
    1       &     0 &      0 &        0 &        0 \\
         0  &  0.87 &   0.02 &        0 &        0 \\
         0  &  0.09 &   1.09 &        0 &        0 \\
         0  &       0 &        0 &   0.86 &  -0.01 \\
         0  &       0 &        0 &  -0.03 &   0.98
\end{array}
\right) \;.
\eea
Our conventions are such that 
\be\
R^{\rm BSM}({2\,\GeV}) = U^{\msbar}(2\,{\rm GeV}, 3\,{\rm GeV}) R^{\rm BSM}(3\,\GeV) \,.
\nn
\\
\ee\
%--------------------------------------------------------------------------------
\section*{Acknowledgements}
%{\bf Acknowledgements.}
We thank our colleagues of the RBC and UKQCD collaborations.
We are grateful to F.~Mescia and S.~Sharpe for discussions. 
N.~G. and R.~H are supported by the STFC grant ST/G000522/1 and 
acknowledge the EU grant 238353 (STRONGnet).
%\newpage
\bibliography{biblio}{}
\bibliographystyle{h-elsevier}

\end{document}